\documentclass{article}

\usepackage{main}

\begin{document}

\title{Canonicity for Cost-Aware Logical Framework \\ via Synthetic Tait Computability}
\author{Runming Li \\ Carnegie Mellon University \\ runmingl@cs.cmu.edu 
    \and Robert Harper \\ Carnegie Mellon University \\ rwh@cs.cmu.edu
}
\date{\monthyeardate\today}

\maketitle

\begin{abstract}
  In the original work on the cost-aware logical framework by \citet{niu-sterling-grodin-harper>2022}, a dependent variant of the call-by-push-value language for cost analysis, the authors conjectured that the canonicity property of the type theory can be succinctly proved via Sterling's synthetic Tait computability. This work resolves the conjecture affirmatively.
\end{abstract}


\section{Introduction}\label{sec:intro}

Cost-aware logical framework (\calf{}) \cite{niu-sterling-grodin-harper>2022} is an effectful dependent type theory designed for synthetic cost analysis. The type theory is based on call-by-push-value \cite{levy>2003} that distinguishes pure values and effectful computations, extended with dependent types inspired by \citet{vakar>thesis,pedrot-tabareau>2019}.

In \calf{} cost is treated abstractly as a print-like effect: we write $\mstept{X}{c}{e}$ that records $c$ units of abstract cost before running the computation $e : X$. For example, as verified in \citet{niu-sterling-grodin-harper>2022}, internally in \calf{}, one can prove that merge-sort has a time complexity of $O(n \log n)$ by inhabiting the following type:
\[
  l : \listty{\nat} \vdash \textsf{msort}~l =_{\F{\listty{\nat}}} \mstept{\F{\listty{\nat}}}{\len{l}\log{\len{l}}}{\ret{\textsf{sort}~l}}
\]
where $\textsf{msort} : \listty{\nat} \pto \F{\listty{\nat}}$ is the merge-sort function instrumented with step counting effect, and $\textsf{sort}$ is the specification/value level sorting function.

One of the important contributions of \calf{} is a \emph{synthetic phase distinction} between cost and behavior of programs. Originally invented in the context of module calculus \cite{harper-mitchell-moggi>1989}, the phase distinction is a way to separate static (kinds and types) components and dynamic (terms) components of a module, where dynamic components can depend on static components but not vice versa. \citet{sterling-harper>2021} extended the static/dynamic phase distinction to a synthetic setting, where a type-theoretical proposition representing the \emph{static} phase can be used to control the phase distinction. When the proposition representing the static phase is inhabited, the dynamic components are trivialized, achieving the non-interference between static and dynamic components. Similarly, \calf{} observed that cost analysis displays a similar structure: the cost of a program possibly depends on the behavior of the program, but behavior of program can be analyzed independent of how much it costs. Crucially, \calf{} postulates a new phase $\beh$ for \emph{behavioral} phase in which the cost is trivialized. Under the behavioral phase, only program behaviors are considered, so one can prove that, for example, merge-sort is equal to insertion-sort because they both sort the list:
\[
  \_ : \beh \vdash \textsf{msort} =_{\listty{\nat} \pto \F{\listty{\nat}}} \textsf{isort}.
\]
This is achieved by considering an axiom $\mathsf{step}_\beh$ which states that all step counting instrumentations are silent under the behavioral phase:  
\[
    \mathsf{step}_\beh : \beh \to \mstept{X}{c}{e} = e.
\]
However, in the cost-sensitive phase, merge-sort and insertion-sort, each instrumented with cost step counting (for example, one unit of cost for each comparison operation), are different because they have different cost.

The original work of \calf{} provides a categorical model of the type theory following the standard Eilenberg-Moore model of call-by-push-value in which computation types are interpreted as algebras over a writer monad that keep track of costs. Another important meta-theoretical property of \calf{} as a programming language is the canonicity property: roughly it states that every computation is equal to incurring some amount of cost and then returning a value. The authors of \calf{} conjectured that the canonicity property holds for \calf{} and can be proved succinctly via Sterling's synthetic Tait computability \cite{sterling>thesis}. Formally the theorem is stated as follows:

\begin{theorem}[Canonicity]
  \label{thm:canonicity}
  For every closed computation $e : \F{\nat}$, there exists $c : \costty$ and $n : \mathbb{N}$ such that $e = \mstep{c}{\ret{\num{n}}}$, where $\textsf{suc}^n$ is the $n$-fold successor.
\end{theorem}

This work proves the theorem using synthetic Tait computability, thereby resolving the conjecture from the original development of \calf{} affirmatively. We begin by recalling some tools for \emph{synthetic phase distinctions} and techniques of \emph{synthetic Tait computability}.

\subsection{Toolbox for synthetic phase distinctions}

In this section we collect some useful type structures that are used in the synthetic phase distinctions.

\subsubsection{Propositions}
As mentioned earlier, synthetically we can control the phase distinctions by an intuitionistic proposition. A proposition is a type $\varphi$ such that any two inhabitants of $\varphi$ are equal, that is, a type that satisfies the property $\textsf{isProp}(\varphi) \isdef (a, b : \varphi) \to a = b$. Semantically each proposition is interpreted as a distinguished subterminal object. In \calf{}, we use $\beh$ to represent the behavioral/cost phase distinction; in synthetic Tait computability, we use $\syn$ to represent the syntax/semantics phase distinction.

\subsubsection{Extension types}
Originally introduced in the context of homotopy type theory \cite{riehl-shulman>2017}, extension types are a way to restrict elements of a certain type under a proposition. For a proposition $\varphi$ and an element $a_0 : \varphi \to A$, the extension type $\extTy{A}{z}{\varphi}{a_0}$ classifies all elements $a : A$ that are equal to $a_0$ under the proposition $\varphi$. This is a generalization of singleton types \cite{stone>thesis} to a phased setting; indeed, when $\varphi$ holds, the extension type $\extTy{A}{z}{\varphi}{a_0}$ contains exactly one element $a_0$. 

Already in the context of synthetic phase distinctions, extension types play many roles. In \citet{sterling-harper>2021} it is used to model sharing specification in an ML module calculus. In \citet{grodin-li-harper>2025} it is used to give behavioral specifications of algorithms and data structures. Last but not least, in synthetic Tait computability \cite{sterling>thesis} it is used to restrict the syntax component of a computability structure, which will be the main use in this work.

Notationally, we leave the coercion between the extension type and the original type implicit: if $a : \extTy{A}{z}{\varphi}{a_0}$ then we write $a : A$ as well.

\subsubsection{Open modality}
For a proposition $\varphi$, the open modality $\open_{\varphi}A \isdef \varphi \to A$ is the lex idempotent reader monad for $\varphi$ that enters the phase corresponding to $\varphi$. We write $\eta_{\open_{\varphi}} : A \to \open_{\varphi}A \isdef \lambda (a : A). \lambda (\_ : \varphi). a$ for the monadic return operation. When $\eta_{\open_{\varphi}}: A \to \open_{\varphi}A$ is an isomorphism, then we say that $A$ is $\open_{\varphi}$-modal.\footnote{In other work this notion has been referred to as open-modal/$\Op$-modal~\cite{rijke-shulman-spitters>2020}, $\varphi$-transparent~\cite{sterling-harper>2022,sterling>2022-logical-relations}, purely extensional types~\cite{niu-sterling-grodin-harper>2022,grodin-niu-sterling-harper>2024}, and behavioral types~\cite{grodin-li-harper>2025}.} Open modality induces an open subuniverse $\open_{\varphi} \univ \isdef \varphi \to \univ$ that classifies all $\open_{\varphi}$-modal types. To be precise, the collection of all types $A : \open_{\varphi} \univ$ can be given via the decoding function
\begin{align*}
  \textsf{El}_{\open_{\varphi} \univ}     & : \open_{\varphi} \univ \to \univ \\
  \textsf{El}_{\open_{\varphi} \univ} (A) & \isdef (z : \varphi) \to A~z.
\end{align*}

\subsubsection{Closed modality}
For a proposition $\varphi$, the closed modality $\closed_{\varphi}A$ is a lex idempotent monad defined as the pushout of $\pi_1 : A \times \varphi \to A$ along the map $\pi_2 : A \times \varphi \to \varphi$, or equivalently a quotient type that equates all elements under the phase as follows:
\begin{center}
  \begin{minipage}{0.5\linewidth}
    \[\begin{tikzcd} {A \times \varphi} & \varphi \\
        A & {\closed_{\varphi} A}
        \arrow["{\pi_1}", from=1-1, to=2-1]
        \arrow["{\pi_2}", from=1-1, to=1-2]
        \arrow["{\star_{\varphi}}", from=1-2, to=2-2]
        \arrow["{\eta_{\closed_{\varphi}}}", from=2-1, to=2-2]
        \arrow["\lrcorner"{anchor=center, pos=0.125, rotate=180}, draw=none, from=2-2, to=1-1]
      \end{tikzcd}\]
  \end{minipage}%
  \begin{minipage}{0.5\linewidth}
    \iblock{
      \mhang{\kw{data}~\closed_{\varphi}~(A : \univ) : \univ~\kw{where}}{
        \mrow{\Label{\eta_{\closed_{\varphi}}} : A \to \closed_{\varphi} A}
        \mrow{\Label{\star_{\varphi}} : \{\_ : \varphi\} \to \closed_{\varphi} A}
        \mrow{\Label{law} : (a : A)~\{\_ : \varphi\} \to \Label{\eta_{\closed_{\varphi}}} a = \Label{\star_{\varphi}}}
      }
    }
  \end{minipage}
\end{center}

The quotient case $\Label{law}$ induced by the pushout must be respected by the elimination form of the closed modality. We may succinctly gives the induction principle using extension types, following \citet{sterling>thesis}:
\begin{mathpar}
  \inferrule[$\closed$-Ind]
  {e : \closed_{\varphi} A
    \qquad
    B : \closed_{\varphi} A \to \univ
    \qquad
    f : (a : A) \to \extTy{B(\eta_{\closed_\varphi}(a))}{z}{\varphi}{g}
    \qquad
    g : \varphi \to B(\star_\varphi)}
  {\try{e}{f}{\varphi}{g} : \extTy{B(e)}{z}{\varphi}{g}}.
\end{mathpar}

Similarly, non-dependent elimination form can be given as follows:
\begin{mathpar}
    \inferrule[$\closed$-Rec]
    {e : \closed_{\varphi} A
      \qquad
      f : A \to \extTy{B}{z}{\varphi}{g}
      \qquad
      g : \varphi \to B}
    {\try{e}{f}{\varphi}{g} : \extTy{B}{z}{\varphi}{g}}.
  \end{mathpar}

The corresponding $\beta$ and $\eta$ laws follows from the universal property of the pushout. We write $\mapp_{\closed{\varphi}} : (A \to B) \to \closed_{\varphi}A \to \closed_{\varphi}B$ for the functorial action of the closed modality.

When $\eta_{\closed_{\varphi}} : A \to \closed_{\varphi} A$ is an isomorphism, then we say that $A$ is $\closed_{\varphi}$-modal.\footnote{In other work this notion has been referred to as closed-modal/$\Cl$-modal~\cite{rijke-shulman-spitters>2020}, $\Op$-connected~\cite{rijke-shulman-spitters>2020}, $\varphi$-sealed~\cite{sterling-harper>2021-metalanguage-multi-phase-modularity,sterling-harper>2022,sterling>2022-logical-relations}, purely intensional~\cite{niu-sterling-grodin-harper>2022,grodin-niu-sterling-harper>2024} types, and algorithmic types \cite{grodin-li-harper>2025}.} In particular it means $A$ is trivialized under the phase: $\open_{\varphi}A \cong \unit$ when $A$ is $\closed_{\varphi}$-modal. Immediately the extension type $\extTy{A}{z}{\varphi}{a_0}$ is $\closed_{\varphi}$-modal because under $\varphi$ it is a singleton type.

\subsubsection{Strict glue types}
An important property of synthetic phase distinctions is that the closed parts of a type can \emph{depend} on the open parts but not vice versa. This fact can be captured via a $\Sigma$-like type called the strict glue type \cite{sterling-harper>2022,yang>thesis}.
\begin{mathpar}
  \inferrule[Glue]
  {A : \open_{\varphi}\univ
    \qquad
    B : \textsf{El}_{\open_{\varphi} \univ}(A) \to \univ
    \qquad
    a : \textsf{El}_{\open_{\varphi} \univ}(A) \vdash B~a \text{ is } \closed_{\varphi}\text{-modal}
  }
  {\glue{x}{A}{B~x} : \extTy{\univ}{z}{\varphi}{A}}
\end{mathpar}
In particular, the first component of the strict glue type is $\open_{\varphi}$-modal and the second component is $\closed_{\varphi}$-modal. The extension type ensures that, under $\varphi$, $\glue{x}{A}{B~x} = A$ strictly, because under $\varphi$ a $\closed_{\varphi}$-modal type is trivialized. The elements of the strict glue type are pairs $\glueEx{\varphi}{a}{b}$ where $a : A$ and $b : B~a$. Moreover, under $\varphi$, $\glueEx{\varphi}{a}{b} = a$ strictly. We write $\pi_\circ$ and $\pi_\bullet$ for the first and second projections respectively.

Importantly, every type has an open and a closed component via the fracture theorem \cite{rijke-shulman-spitters>2020}; that is, every type is isomorphic to a strict glue type. \citet{yang>thesis} gives this isomorphism explicitly. \citet{grodin-li-harper>2025} considers a slightly different, but equivalent, glue type (there fibered, here indexed) in the context of behavioral/cost phase distinction of \calf{} and shows that it is useful for combining the behavior and cost of a data structure. In this work, we will mostly use the strict glue type to model the syntax/semantics phase distinction in the context of synthetic Tait computability.

The strict glue type can be defined in terms of the realignment/strictification axiom \cite{birkedal-bizjak-clouston-grathwohl-spitters-vezzosi>2016,orton-pitts>2016,sterling-harper>2022,sterling>thesis} that turns a partial isomorphism under $\varphi$ into a strict equality. In such case the strict glue type can be defined as the realigned $\Sigma$ type of the same data $A$ and $B$: $$\glue{x}{A}{B~x} \isdef \textsf{realign } (x : (z : \varphi) \to A~z) \times B~x.$$ 
\citet{gratzer-shulman-sterling>2024} shows that every Grothendieck topos has a cumulative hierarchy of universes satisfying the realignment axiom. We will give an explicit interpretation of the strict glue type in the particular category relevant to this work in \cref{sec:interpretation}.

\subsection{Synthetic Tait computability}
Syntactic meta-properties of type theories can be established using the method of logical relations \cite{tait>1967, plotkin>1973}, in which each type is equipped with a hereditarily defined computability predicate. \citet{kaposi-huber-sattler>2019} showed that categorical gluing \cite[\S 4.10]{crole>1994}, the construction of a category by gluing two categories along a functor, corresponds to proof-relevant logical relations. In particular, gluing along the global section functor yields canonicity for a type theory. This technique has been applied to various canonicity and normalization results, including simply-typed $\lambda$-calculus \cite{fiore>2002, sterling-spitters>2018}, System F \cite{altenkirch-hofmann-streicher>1996}, and dependent type theory \cite{coquand>2019}.

Sterling introduced the method of synthetic Tait computability \cite{sterling>thesis}, which replaces explicit reasoning in glued categories with reasoning in the internal languages of the Artin gluing of topoi. These internal languages often support rich dependent type structures, allowing certain categorical constructions to be reduced to type-theoretic programming. The method was first applied by \citet{sterling-harper>2021} to prove parametricity for an ML module calculus and has since become a key tool for proving metatheorems in dependent type theory. Notable applications include normalization for Cartesian cubical type theory \cite{sterling-angiuli>2021} and multimodal type theory \cite{gratzer>2022}. For a pedagogical introduction, see \citet{sterling>2022-logical-relations}, as well as dissertations on the subject \cite{sterling>thesis, gratzer>thesis, yang>thesis}.

\subsection{Call-by-push-value}
Levy's call-by-push-value \cite{levy>2003} is a polarized $\lambda$-calculus that separates the type structures into pure value types and effectful computation types, following the slogan \emph{values are, computations do}. Call-by-push-value features a free-forgetful adjunction between value types and computation types $\F \dashv \UU$. The induced monad $\UU\F$ is a natural candidate for modeling computational effects. The fundamental operations governs the type structures are as follows (writing $A$ for value types and $X$ for computation types): 
\begin{mathpar}
    \inferrule[$\F$-Intro]
    {a : A}
    {\ret{a} : \F{A}}

    \inferrule[$\F$-Elim]
    {e_0 : \F{A} \\ x : A \vdash e_1 : X}
    {\bind{e_0}{x.e_1} : X}

    \inferrule[$\UU$-Intro]
    {e : X}
    {\thunk{e} : \UU{X}}

    \inferrule[$\UU$-Elim]
    {a : \UU{X}}
    {\force{a} : X}
\end{mathpar}
where $\thunk{-}$ and $\force{-}$ are a pair of isomorphisms.
\calf{} is naturally situated inside call-by-push-value, since cost counting is treated as an abstract computational effect. We refer readers to \citet{levy>2003} for a detailed introduction to call-by-push-value and \citet{niu-sterling-grodin-harper>2022} for an introduction to the cost-aware variant of call-by-push-value.

\paragraph{Synopsis}
This paper is organized as follows.
In \cref{sec:calf} we recall the cost-aware logical framework \calf{} by giving its signature in a logical framework, which gives a category of judgments that represents the syntax of the type theory.
In \cref{sec:computability} we give explicitly the computability structure of \calf{} according to the techniques of synthetic Tait computability.
In \cref{sec:interpretation} we interpret the computability structure of \calf{} into a glued category that represents the semantics of the type theory, which allows us to conclude the canonicity property.
Finally, in \cref{sec:conclusion}, we discuss related work and future directions.

\section{A refresher on cost-aware logical framework}\label{sec:calf}

The presentation of a type theory can be given succinctly as a signature in a logical framework following the \emph{judgments as types} slogan \cite{harper-honsell-plotkin>1993,sterling>thesis}. Following the original development of \calf{}, we give the signature of \calf{} in a semantic logical framework \cite{harper>2021,uemura>thesis,yang>2025,sterling>thesis}, that is, the internal language of locally Cartesian closed categories. The adequacy of such presentation follows from \citet{gratzer-sterling>2021} on defining dependent type theories in locally Cartesian closed categories. 

In particular, we assume a universe of \emph{judgments} $\jdg$ closed under $\unit$, $\Pi$, $\Sigma$, and extensional equality types, in which we declare judgments of the type theory as constants. Binding and scope are handled by the framework-level $\Pi$ types via higher-order abstract syntax. The equational theory of the object language is given by framework-level extensional equality between constants. We write $(\alpha,\beta) : A \cong B$ when $\alpha$ and $\beta$ are the forward map $A \to B$ and backward map $B \to A$ of an isomorphism $A \cong B$. We write framework-level $\Pi$ and $\Sigma$ types using Nuprl style \cite{constable>1986} notation $(x : A) \to B~x$ (with introduction form as $\lambda$ and elimination form as juxtaposition) and $(x : A) \times B~x$ (with introduction form as a pair and elimination form as projections) respectively. 

We fix a monoid $\costty$ with a pre-order and its associated monoid operations $(\costty, 0, +, \le)$ for the signature of \calf{}. In particular \calf{} features a dichotomy of value types $\tpv$ and computation types $\tpc$ in the manner of call-by-push-value and a pair of adjoint type constructors $\F$ and $\UU$ that mediates between value types and computation types. Following the original development of \calf{}, a term of a computation type $X$ is defined as a term of value type $\UU{X}$, resulting in a version of call-by-push-value where the isomorphisms $\thunk{-}$ and $\force{-}$ are suppressed. To keep track of cost counting, \calf{} includes a $\mstep{c}{e}$ instruction on computations, and a $\beh$avioral phase (with its corresponding open/closed modality) that synthetically ``erases'' the cost counting. 

The signature for \calf{} is given as follows:
{\allowdisplaybreaks
\begin{align*}
  \kw{record}~                     & \Sigma_{\calf}~\kw{where}                                                                 \\
  \tpv                             & : \jdg                                                                                    \\
  \tpc                             & : \jdg                                                                                    \\
  \mathsf{tm}^+                    & : \tpv \to \jdg                                                                           \\
  \mathsf{U}                       & : \tpc \to \tpv                                                                           \\
  \mathsf{F}                       & : \tpv \to \tpc                                                                           \\
  \mathsf{tm}^{\ominus}(X)         & \coloneqq \tmv{\UU{X}}                                                                    \\
  \mathsf{ret}                     & : \impl{\isof{A}{\tpv}} (\isof{a}{\tmv{A}}) \to \tmc{\F{A}}                               \\
  \mathsf{bind}                    & : \impl{\isof{A}{\tpv}, \isof{X}{\tpc}} \tmc{\F{A}} \to (\tmv{A} \to \tmc{X}) \to \tmc{X} \\
  \mathsf{bind}_{\beta}            & : \impl{A, X, a, f} \bind{\ret{a}}{f} = f(a)                                              \\
  \mathsf{bind}_{\eta}             & : \impl{A, e} \bind{e}{\mathsf{ret}} = e                                                  \\
  \mathsf{bind}_{\mathsf{assoc}}   & : \impl{A, X, e, f, g} \bind{(\bind{e}{f})}{g} = \bind{e}{(\lambda a.\, \bind{f(a)}{g})}  \\
  \\
  \stepp                    & : \impl{X} \mathbb{C} \to \tmc{X} \to \tmc{X}                                             \\
  \stepp_{0}                & : \impl{X, e} \mstep{0}{e} = e                                                            \\
  \stepp_{+}                & : \impl{c_1,c_2, X, e} \mstep{c_1}{{\mstep{c_2}{e}}} = \mstep{c_1 + c_2}{e}             \\
  \\
  \beh                             & : \jdg                                                                                    \\
  \beh_{\mathsf{uni}}              & : \impl{\isof{u,v}{\beh}} u = v                                                           \\
  \open_{\beh}(\mathcal{J})        & \coloneqq \beh \to \mathcal{J}                                                            \\
  \stepp_{\beh}             & : \impl{c, X, e} \open_{\beh}(\mstep{c}{e} = e)                                           \\
  \\
  \Op^+                                   & : \tpv \to \tpv \\
  (\app_{\Op}, \lambdap_{\Op})              & : \impl{A} \tmv{\Op^+(A)} \cong \Op_{\beh}(\tmv{A}) \\
  \\
  \Cl^+                            & : \tpv \to \tpv           \\
  \eta_{\Cl}                     & : \impl{A} \tmv{A} \to \tmv{\Cl^+(A)} \\
  \star                            & : \impl{A} \beh \to \tmv{\Cl^+{A}} \\
  \mathsf{law}_{\Cl}             & : \impl{A, a, z} \eta_{\Cl}(a) = \star(z) \\
  \mathsf{ind}_{\Cl}             & : \impl{A} (\isof{a}{\tmv{\Cl^+(A)}}) \to (X : \tmv{\Cl^+(A)} \to \tpc) \to \\
                                   & (\isof{x_0}{(\isof{a}{\tmv{A}}) \to \tmc{X(\eta_{\Cl}(a))}}) \to \\
                                   & (\isof{x_1}{(\isof{z}{\beh}) \to \tmc{X(\star(z))}}) \to \\
                                   & ((\isof{a}{\tmv{A}}) \to (\isof{z}{\beh}) \to x_0(a) = x_1(z)) \to \\
                                   & \tmc{X(a)} \\
  {\mathsf{ind}_{\Cl}}_{\eta_{\Cl}}             & : \impl{A, a, x_0, x_1, h} \mathsf{ind}_{\Cl}(\eta_{\Cl}(a); X; x_0; x_1; h) = x_0(a) \\
  {\mathsf{ind}_{\Cl}}_{\star}             & : \impl{A, z, x_0, x_1, h} \mathsf{ind}_{\Cl}(\star(z); X; x_0; x_1; h) = x_1(z) \\
  \\
  \Pi                              & : (\isof{A}{\tpv}, \isof{X}{\tmv{A} \to \tpc}) \to \tpc                                   \\
  (\mathsf{ap}, \mathsf{lam})      & : \impl{A,X} \tmc{\Pi(A; X)} \cong (\isof{a}{\tmv{A}}) \to \tmc{X(a)}                     \\
  \Sigma                           & : (\isof{A}{\tpv}, \isof{B}{\tmv{A} \to \tpv}) \to \tpv                                   \\
  (\mathsf{unpair}, \mathsf{pair}) & : \impl{A,B} \tmv{\signeg{A}{B}} \cong (a : \tmv{A}) \times \tmv{B(a)}                    \\
  \\
  \mathsf{eq}                      & : (\isof{A}{\tpv}) \to \tmv{A} \to \tmv{A} \to \tpv                                       \\
  (\mathsf{ref}, \mathsf{self})    & : \impl{A} (\isof{a,b}{\tmv{A}}) \to
  (a =_{\tmv{A}} b) \cong \tmv{\eqty{A}{a}{b}}\\
  \\
  \nat                             & : \tpv                                                                                    \\
  \zero                            & : \tmv{\nat}                                                                              \\
  \mathsf{suc}                     & : \tmv{\nat} \to \tmv{\nat}                                                               \\
  \mathsf{ind}                     & : (\isof{n}{\tmv{\nat}}) \to (\isof{X}{\tmv{\nat} \to \tpc}) \to \tmc{X(\zero)} \to       \\
                                   & ((\isof{n}{\tmv{\nat}}) \to \tmc{X(n)} \to \tmc{X(\suc{n})}) \to \tmc{X(n)}               \\
  \mathsf{ind}_{\zero}             & : \impl{X, e_0, e_1} \mathsf{ind}(\zero; X; e_0; e_1) = e_0                                  \\
  \mathsf{ind}_{\textsf{suc}}              & : \impl{X, n, e_0, e_1} \mathsf{ind}(\suc{n}; X; e_0; e_1) = e_1 (n) (\ind{n}{X}{e_0}{e_1})         \\
  \\
  \mathsf{ap}_{\stepp}      & : \impl{A,X,f,a,c} \ap{\mstep{c}{f}}{a} = \mstep{c}{\ap{f}{a}}                            \\
  \mathsf{bind}_{\stepp}    & : \impl{A,X,e,f,c} \bind{\mstep{c}{e}}{f} = \mstep{c}{\bind{e}{f}}
\end{align*}
}

We slightly deviate from the original development of \calf{} in the following ways:
\begin{enumerate}
    \item We remove the mixed polarized Sigma type $\Sigma^{+-} : (\isof{A}{\tpv}, \isof{X}{\tmv{A} \to \tpc}) \to \tpc$ and its associated equations such as $\mathsf{pair}_{\mathsf{step}} : \mstep{c}{(a,e)} = (a , \mstep{c}{e})$. This is because such type allows for a projection elimination rule that is able to ``forget'' effects: specifically, taking the first projection of a computation $\mstep{c}{(a,e)}$ gives $a : A$ which is a pure value, effectively discarding the cost $c$. This is not desirable in a cost-accounting setting, where inadvertent omission of cost can lead to unsoundness of cost analysis. To recover computation level $\Sigma$ types, the solution is to consider a dependent copower with a ``split'' elimination in the style of Enriched Effect Calculus \cite{egger-mogelberg-simpson>2014}. However, such elimination requires a linear typing context, which is unclear how to model in the logical framework of locally Cartesian closed categories. We leave further development of this idea to future work.
    \item We consider a purely value level equality reflection by requiring $\textsf{ref}$ and $\textsf{self}$ to be isomorphisms, and remove the need to discuss computation level equality reflection. This is consistent with the later development of \citet{grodin-niu-sterling-harper>2024}, and significantly simplifies the computability structure of equality types which we will discuss in \cref{sec:computability}.
    \item We replace the $\mathsf{lam}_{\stepp} : \lam{\lambda x. \mstep{c}{f~x}} = \mstep{c}{\lam{f}}$ equation with $\mathsf{ap}_{\stepp} : \ap{\mstep{c}{f}}{a} = \mstep{c}{\ap{f}{a}}$ which is more natural in the context of call-by-push-value, and essential in the later development of canonicity proof. The original equation $\mathsf{lam}_{\stepp}$ is derivable from $\mathsf{ap}_{\stepp}$, but not vice versa.
\end{enumerate}

\subsection{Category of judgments}

A signature defined in the logical framework induces a category of judgments $\mathcal{C}$ \cite{sterling>thesis,yang>2025} where the objects are judgments $\mathcal{J} : \jdg$ and morphisms are equivalence class of deductions (modulo judgmental equality) between judgments $f : \mathcal{J}_1 \to \mathcal{J}_2$. For example, judgments of the form 
\[
    x_1 : A_1, x_2 : A_2, \cdots x_n : A_n \vdash e : B 
\]
correspond to morphisms 
\[
    e : \tmv{A_1} \times \tmv{A_2} \times \cdots \times \tmv{A_n} \to \tmv{B}.
\]
In particular, a closed term $a$ of type $A$ is then a morphism $a : \terminal_{\mathcal{C}} \to \tmv{A}$.

Since the universe $\jdg$ is closed under $\unit$, $\Pi$, $\Sigma$, and extensional equality types, the category of judgments $\mathcal{C}$ is a locally Cartesian closed category whose internal language is exactly extensional Martin-L\"of type theory extended with the constants declared in the signature $\Sigma_\calf{}$. 

Therefore, a model of \calf{} in a locally Cartesian closed category $\mathcal{E}$ can be viewed in the following two ways:
\begin{enumerate}
    \item A locally Cartesian closed functor $\mathcal{C} \to \mathcal{E}$ in the style of functorial semantics \cite{lawvere>thesis}.
    \item A term of type $\Sigma_{\calf}$ (viewed as a dependent record, or an iterated $\Sigma$ type) in an extensional dependent type theory (the internal language of $\mathcal{E}$) where $\jdg$ is interpreted as a universe in $\mathcal{E}$. 
\end{enumerate}

The canonicity argument will eventually come down to the first view of functorial models, but usually such a functor is tedious to construct directly. Instead, in the next section we will give a model of \calf{} through the second view in the language of synthetic Tait computability \cite{sterling>thesis} that gives a computability structure for which a canonicity argument can be made. The two views of models being equivalent goes back to \citet{seely>1984}; \citet{yang>2025} proved the same result in the specific case where $\mathcal{C}$ is the category of judgments as described here.

\section{Computability structure of \calf{}}\label{sec:computability}

We axiomatize the language of synthetic Tait computability, a variant of Martin L\"of type theory extended with toolbox of synthetic phase distinctions. We then construct a model of \calf{} in this language in which we can deduce the canonicity property. We will justify these axioms semantically in \cref{sec:interpretation}.

\begin{axiom}
    The language of synthetic Tait computability is an extensional Martin L\"of type theory with a universe hierarchy $\univ_0 : \univ_1 : \univ_2 : \cdots$. For the sake of discussion we assume a pair of universes $\univ_i$ and $\univ_j$ such that $i < j$.
\end{axiom}

\begin{axiom}
    We axiomatize a proposition $\syn$ that represents the syntax/semantics phase distinction. Then we use the toolbox of synthetic phase distinctions to obtain useful type structures such as extension types $\extTy{A}{z}{\syn}{a_0}$, the open modality $\open_{\syn}A$, the closed modality $\closed_{\syn}A$, and strict glue types $\glue{x}{A}{B~x}$.
\end{axiom}

Assume we have a structure $\red{\M} : \Sigma_{\calf{}}$ where $\jdg$ is interpreted as the $\syn$-open subuniverse $\open_{\syn}\univ$. This structure represents a generic \emph{syntactic} model of \calf{}. The use of open subuniverse makes sure that we can only talk about the syntax of the type theory under the syntactic phase when we have a proof of $\syn$.

In this section, we show how to construct a semantic computability structure $\blue{\M} : \extTy{\Sigma_{\calf{}}}{z}{\syn}{\red{\M}}$ of a sufficiently large universe $\univ_j$ that restricts to the syntactic model $\red{\M}$ under the syntactic phase. From this model we will be able to prove a fundamental theorem of synthetic Tait computability, analogous to the fundamental theorem of traditional logical relations.

Viewing $\Sigma_{\calf{}}$ as a dependent record, we can take projections such as $\red{\M}.\red{\tpv} : \open_{\syn}\univ$ and $\red{\M}.\red{\tmvp} : \red{\M}.\red{\tpv} \to \open_{\syn}\univ$. This means we are to define components of $\blue{\M}$ that restricted to their syntactic counterparts under the syntactic phase. In other words, we need to define $\blue{\M}.\blue{\tpv} : \extTy{\univ_j}{z}{\syn}{\red{\M}.\red{\tpv}}$, $\blue{\M}.\blue{\tmvp} : \extTy{\blue{\M}.\blue{\tpv} \to \univ_j}{z}{\syn}{\red{\M}.\red{\tmvp}}$, and so on. Notationally, we use \red{red} for the syntax and \blue{blue} for the semantics, so we may drop ``$\red{\M}.$'' and ``$\blue{\M}.$'' for brevity.

\begin{axiom}
    \label{ax:beh}
    We axiomatize a proposition $\beh$ that represents the behavioral/cost phase distinction of \calf{}. Furthermore we require that $\beh$ restricts to the actual syntactic $\red{\beh}$ under the syntactic phase, that is, $\beh : \extTy{\univ_j}{z}{\syn}{\red{\beh}}$. Again we have available its corresponding type structures such as the closed modality $\closed_{\beh}A$ from the toolbox.
\end{axiom}

Now we are ready to define the computability structure $\blue{\M} : \extTy{\Sigma_{\calf{}}}{z}{\syn}{\red{\M}}$.

\subsection{Computability structure of judgments}
\subsubsection{Value types}
Value types in call-by-push-values are exactly like the ordinary types in Martin-L\"of type theory. Following \citet{sterling-harper>2021}, we define the computability structure of value types as the collection of all terms of a syntactic type $A$: 
\iblock{
    \mrow{\Label{\blue{\M}.\blue{\tpv}} : \extTy{\univ_j}{z}{\syn}{\red{\tpv}}}
    \mrow{\Label{\blue{\M}.\blue{\tpv}} = \glue{A}{\red{\tpv}}{\extTy{\univ_i}{z}{\syn}{\red{\tmvp}~A}}}
    \mrow{}
    \mrow{\Label{\blue{\M}.\blue{\tmvp}} : \extTy{\blue{\tpv} \to \univ_j}{z}{\syn}{\red{\tmvp}}}
    \mrow{\Label{\blue{\M}.\blue{\tmvp}} A = \pi_\bullet A}
}

\subsubsection{Computation types}
The categorical semantics of \calf{}, an instance of the standard Eilenberg-Moore model of call-by-push-value, tells us that the semantics of a computation type requires more structure than value types. In particular, a computation type $X : \tpc$ is interpreted as an algebra over the writer monad $\costty \times -$. This suggests that the computability structure of a computation type $X$ should likewise be an algebra over monad. 

We define a cost algebra structure $\alg$ as a record with a carrier $\Label{C}$ and a mapping function $\Label{sstep}$ (stands for \textit{s}emantic step) corresponding to the writer monad $\costty \times -$ as shown below. The mapping function is required to satisfy the usual algebraic laws of a monad. Indeed $\Label{sstep_0}$ and $\Label{sstep_+}$ are the unit and multiplication laws of monad algebra. Ordinary call-by-push-value would have exactly those first four components, but we add an additional equation $\Label{sstep_{\beh}}$ in the cost algebra that reflects a special property of erasing cost under the behavioral phase in \calf{}, in anticipation of having to satisfy the equation $\mathsf{step}_\beh$. Importantly the syntactic $\red{\stepp}$ and its associated equations forms exactly such an algebra $\Label{SynAlg}$. 

\newpage
\iblock{
    \mhang{\kw{record}~\alg~\kw{where}}{
        \mrow{\Label{C} : \univ_i}
        \mrow{\Label{sstep} : \costty \to \Label{C} \to \Label{C}}
        \mrow{\Label{sstep_{0}} : \Label{sstep}~0~x = x}
        \mrow{\Label{sstep_{+}} : \Label{sstep}~c_1~(\Label{sstep}~c_2~x) = \Label{sstep}~(c_1 + c_2)~x}
        \mrow{\Label{sstep_{\beh}} : \open_{\beh}(\Label{sstep}~c~x = x)}
    }
    \mrow{}
    \mrow{\Label{SynAlg} : \syn \to (X :\red{\tpc}) \to \alg}
    \mrow{\Label{SynAlg}\proj{C} = \red{\tmcp}~X}
    \mrow{\Label{SynAlg}\proj{sstep} = \red{\stepp}}
    \mrow{\Label{SynAlg}\proj{sstep_{0}} = \red{\stepp_{0}}}
    \mrow{\Label{SynAlg}\proj{sstep_{+}} = \red{\stepp_{+}}}
    \mrow{\Label{SynAlg}\proj{sstep_{\beh}} = \red{\stepp_{\beh}}}
    \mrow{}
    \mrow{\Label{\blue{\M}.\blue{\tpc}} : \extTy{\univ_j}{z}{\syn}{\red{\tpc}}}
    \mrow{\Label{\blue{\M}.\blue{\tpc}} = \glue{X}{\red{\tpc}}{\extTy{\alg}{z}{\syn}{\textsf{SynAlg}~\_~X}}}
}

\subsection{Computability structure of types and terms}

\subsubsection{$\UU$ and $\F$ adjunction}

The usual categorical semantics of semantics suggests that $\F \dashv \UU$ is a free-forgetful adjunction. This means that the computability structure of $\F$ should likewise be a free cost algebra and the computability structure of $\UU$ should be a forgetful functor. 

\iblock{
    \mrow{\Label{\blue{\M}.\blue{\UU}} : \extTy{\blue{\tpc} \to \blue{\tpv}}{z}{\syn}{\red{\UU}}}
    \mrow{\Label{\blue{\M}.\blue{\UU}} X = \glueEx{\syn}{\red{\UU}X}{(\pi_\bullet X)\proj{C}}}
    \mrow{}
    \mrow{\Label{\blue{\M}.\blue{\tmcp}} : \extTy{\blue{\tpc} \to \univ_j}{z}{\syn}{\red{\tmcp}}}
    \mrow{\Label{\blue{\M}.\blue{\tmcp}} X = \blue{\tmvp}(\blue{\UU}X)}
}

It is instructive to check that under the syntactic phase $\Label{\blue{\M}.\blue{\UU}} X = \glueEx{\syn}{\red{\UU}X}{(\pi_\bullet X)\proj{C}}$ restricts to $\red{\UU}X$ by the definition of the strict glue type. Indeed here $\pi_\bullet X$ is a cost algebra for the computation type $X$ and $\proj{C}$ gives the underlying carrier, which corresponds to a forgetful functor in the categorical semantics.

To give a computability structure for $\F$ type, we need to provide a cost algebra structure $\Label{FAlg}$. This is the centerpiece, and arguably the most complex part, of the canonicity proof. We break it down into pieces. 

\iblock{
    \mrow{\Label{\blue{\M}.\blue{\F}} : \extTy{\blue{\tpv} \to \blue{\tpc}}{z}{\syn}{\red{\F}}}
    \mrow{\Label{\blue{\M}.\blue{\F}} A = \glueEx{\syn}{\red{\F}A}{\Label{FAlg}~A}}
}

\iblock{
    \mrow{\Label{FAlg} : (A : \blue{\tpv}) \to \alg}
}

\begin{itemize}
    \item  First, the carrier $\Label{FAlg}\proj{C}$ should contain the key property of the canonicity proof: every computation of type $\F A$ should be $\mstep{c}{\ret{a}}$ for some $a : \tmvp A$ and $c : \costty$. Later specializing the type $A$ to $\nat$ will give us the desired canonicity result, analogous to how the computability predicate at the base type should exactly be the desired property in traditional logical relations arguments. Therefore, $\Label{FAlg}\proj{C}$ consists of a syntactic term of type $\red{\F}A$ and a proof that it is of the form $\mstep{c}{\ret{a}}$. Moreover, in anticipation of needing to satisfy the $\Label{step_{\beh}}$ equation, which erases the cost under $\beh$, we hide the cost $c : \costty$ under $\closed_{\beh}$.
    
    \iblock{
        \mhang{\Label{FAlg}\proj{C} = \glue{e}{\red{\tmcp}(\red{\F}A)}{\closed_{\syn}((a : \blue{\tmvp} A, bc : \closed_{\beh}{\costty}) \times \alpha)}}{
            \mrow{\kw{where}}
            \mrow{\alpha = \try{bc}{\lambda (c : \costty). \open_{\syn}(e = \red{\stepp}^c(\red{\retp}(a)))}{\beh}{\open_{\syn}(e = \red{\retp}(a))}}
        }
    }
    It is important to check that this definition of carrier is properly typed:
    \begin{enumerate}
        \item The second component of the strict glue type is under $\closed_{\syn}$, so it is $\closed_{\syn}$-modal. 
        \item Under $\beh$, the two clauses of the $\textsf{try}$ are equal because $\red{\stepp}^c(\red{\retp}(a)) = \red{\retp}(a)$ by $\red{\stepp_{\beh}}$ equation.
        \item Under $\syn$, $\Label{FAlg}\proj{C}$ is equal to $\Label{SynAlg}\proj{C} = \red{\tmcp}(\red{\F}A)$ by the definition of the strict glue type.
    \end{enumerate}

    \item Second, the role of $\Label{FAlg.sstep}$ is to add $c'$ amount of cost to the semantical component that records cost via $\mapp_{\closed_{\beh}} (c' + \_) (bc)$.
    \iblock{
        \mhang{\Label{FAlg}\proj{sstep} (c' : \costty) ((e, s) : \Label{FAlg.C}) = \glueEx{\syn}{\red{\stepp}^{c'}(e)}{\mapp_{\closed_{\syn}} (\gamma) (s)}}{
            \mrow{\kw{where}}
            \mrow{\gamma : ((a : \blue{\tmvp} A, bc : \closed_{\beh}{\costty}) \times \alpha) \to ((a : \blue{\tmvp} A, bc : \closed_{\beh}{\costty}) \times \alpha)}
            \mrow{\gamma~(a : \blue{\tmvp} A, bc : \closed_{\beh}{\costty}, p : \alpha) = (a, \mapp_{\closed_{\beh}} (c' + \_) (bc), \eta_{\open_{\syn}} (\checkmark))}
        }
    }

    The associated equations are updated with the new cost involving $c'$, and are valid because:
    \begin{enumerate}
        \item $\checkmark: \red{\stepp}^{c'}(e) = \red{\stepp}^{c'+c}(\red{\retp}(a))$ because from $\alpha$ we have an equation $e = \red{\stepp}^c(\red{\retp}(a))$. Then $$\red{\stepp}^{c'}(e) = \red{\stepp}^{c'}(\red{\stepp}^c(\red{\retp}(a))) = \red{\stepp}^{c'+c}(\red{\retp}(a))$$ by the $\red{\stepp_{+}}$ equation.
        \item $\checkmark: \red{\stepp}^{c'}(e) = \red{\retp}(a)$ because again from $\alpha$ we have an equation $e = \red{\retp}(a)$. Then $$\red{\stepp}^{c'}(e) = \red{\stepp}^{c'}(\red{\retp}(a)) = \red{\retp}(a)$$ under the behavioral phase by the $\red{\stepp_{\beh}}$ equation.
    \end{enumerate}

    We further need to check:
    \begin{itemize}
        \item The condition imposed by $\blue{\tpv}$: $\Label{FAlg}\proj{sstep} = \Label{SynAlg}\proj{sstep} = \red{\stepp}$ under $\syn$; this is true because $\glueEx{\syn}{\red{\stepp}^{c'}(e)}{\beta} = \red{\stepp}^{c'}(e)$ under $\syn$ by the definition of the strict glue type.
    \end{itemize}

    \item By examining the definition of $\Label{FAlg}\proj{sstep}$, we can easily see that the equations $\Label{FAlg}\proj{sstep_{0}}$ and $\Label{FAlg}\proj{sstep_{+}}$ holds: the syntactic part of the equations holds by $\red{\stepp_0}$ and $\red{\stepp_+}$ equations, and the semantic part holds by identity of $0$ and associativity of $+$ in the monoid structure $\costty$. 

    \iblock{
        \mrow{\Label{FAlg}\proj{sstep_{0}} = \checkmark}
        \mrow{\Label{FAlg}\proj{sstep_{+}} = \checkmark}
    }
    
    \item Last but not least, to validate $\Label{FAlg}\proj{sstep_{\beh}}$: under $\beh$, we want to show that $$\Label{FAlg}\proj{sstep}~c'~ \glueEx{\syn}{e}{s} = \glueEx{\syn}{e}{s}.$$ Under $\beh$, we have that $\gamma$ is the identity function; therefore $\mapp_{\closed_\syn}(\gamma)(s) = s$, which is what we want.
    
    \iblock{
        \mrow{\Label{FAlg}\proj{sstep_{\beh}} = \eta_{\open_{\beh}} (\checkmark)}
    }
\end{itemize}

\subsubsection{Terms for $\stepp$}
Due to the construction of $\Label{Alg}$, the term and equations for $\blue{\stepp}$ are immediate by projecting from the algebra structure.
\iblock{
    \mrow{\Label{\blue{\M}.\blue{\stepp}} : \extTy{\{X : \blue{\tpv}\} \to \costty \to \blue{\tmcp}(X) \to \blue{\tmvp}(X)}{z}{\syn}{\red{\stepp}}}
    \mrow{\Label{\blue{\M}.\blue{\stepp}} ~\{X\} = (\pi_\bullet X)\proj{sstep}}
    \mrow{}
    \mrow{\Label{\blue{\M}.\blue{\stepp_{0}}} : \extTy{\{X : \blue{\tpv}\} \to \blue{\stepp}^0(e) = e}{z}{\syn}{\red{\stepp_{0}}}}
    \mrow{\Label{\blue{\M}.\blue{\stepp_{0}}} ~\{X\} = (\pi_\bullet X)\proj{sstep_{0}}}
    \mrow{}
    \mrow{\Label{\blue{\M}.\blue{\stepp_{+}}} : \extTy{\{X : \blue{\tpv}\} \to \blue{\stepp}^{c_1}(\blue{\stepp}^{c_2}(e)) = \blue{\stepp}^{c_1 + c_2}(e)}{z}{\syn}{\red{\stepp_{+}}}}
    \mrow{\Label{\blue{\M}.\blue{\stepp_{+}}} ~\{X\} = (\pi_\bullet X)\proj{sstep_{+}}}
    \mrow{}
    \mrow{\Label{\blue{\M}.\blue{\stepp_{\beh}}} : \extTy{\{X : \blue{\tpv}\} \to \open_{\beh}(\blue{\stepp}^c(e) = e)}{z}{\syn}{\red{\stepp_{\beh}}}}
    \mrow{\Label{\blue{\M}.\blue{\stepp_{\beh}}} ~\{X\} = (\pi_\bullet X)\proj{sstep_{\beh}}}
}

\subsubsection{$\bindp$ and $\retp$}
We move on to $\mathsf{bind}$ and $\mathsf{ret}$.

\iblock{
    \mrow{\Label{\blue{\M}.\blue{\retp}} : \extTy{(a : \blue{\tmvp}(A)) \to \blue{\tmcp}(\blue{\F}(A))}{z}{\syn}{\red{\retp}}}
    \mrow{\Label{\blue{\M}.\blue{\retp}}~a = \glueEx{\syn}{\red{\retp}~a}{\eta_{\closed_{\syn}}(a, \eta_{\closed_{\beh}}(0), \eta_{\open_{\syn}}(\checkmark : \red{\retp}(a) = \red{\stepp}^0(\red{\retp}(a))) )}}
}

That is, the semantics of $\retp(a)$ is that it returns the value $a$ with $0$ cost. The equation in \red{red} is true by judgmental equality $\red{\stepp_0}$ of the syntax. 

\iblock{
    \mrow{\Label{\blue{\M}.\blue{\bindp}} : \extTy{\blue{\tmcp}(\blue{\F}(A)) \to ((a : \blue{\tmvp}(A)) \to \blue{\tmcp}(X)) \to \blue{\tmcp}(X)}{}{\syn}{\red{\bindp}}}
    \mhang{\Label{\blue{\M}.\blue{\bindp}}~e~f = \try{\pi_\bullet e}{\lambda (a : \blue{\tmvp}(A), bc : \closed_{\beh}\costty, p : \alpha). \beta}{\syn}{\red{\bindp}~e~f}}{
        \mrow{\kw{where}}
        \mrow{\beta = \try{bc}{\lambda (c : \costty). \blue{\stepp}^c(f~a)}{\beh}{f~a}}
    }
}

That is, the semantics of $\bindp(e, f)$ is that if $e$ incurs cost $c$ and returns $a$, then $\bindp(e, f)$ will apply $f$ to $a$ with $c$ amount of cost added. We need to check that this definition type-checks:
\begin{enumerate}
    \item The inner $\textsf{try}$ clause is valid because under $\beh$, we have $\blue{\stepp}^c(f~a) = f~a$ by the $\blue{\stepp_{\beh}}$ equation above.
    \item The outer $\textsf{try}$ clause is valid because under $\syn$, we have 
    \begin{align*}
        \syn \vdash & \\
        &  \blue{\stepp}^c(f~a) \\
        =~& \red{\stepp}^c(f~a) \tag*{under $\syn$} \\
        =~& \red{\stepp}^c(\red{\bindp}~(\red{\retp}(a))~f) \tag*{by $\red{\bindp_{\beta}}$} \\
        =~& \red{\bindp}~(\red{\stepp}^c(\red{\retp}(a)))~f \tag*{by $\red{\bindp_{\stepp}}$} \\
        =~& \red{\bindp}~e~f \tag*{by $p : \open_\syn (e = \red{\stepp}^c(\red{\retp}(a)))$}
    \end{align*}
    The case where $\beh$ holds is similar. Note that this is analogous to the head expansion lemma in the traditional logical relations proof.  
\end{enumerate}

With these definitions, it is routine calculation to check that the associated equations hold.
\iblock{
    \mrow{\Label{\blue{\M}.\blue{\bindp_\beta}} = \checkmark}
    \mrow{\Label{\blue{\M}.\blue{\bindp_{\eta}}} = \checkmark}
    \mrow{\Label{\blue{\M}.\blue{\bindp_{\textsf{assoc}}}} = \checkmark}
    \mrow{\Label{\blue{\M}.\blue{\bindp_{\stepp}}} = \checkmark}
}

\subsubsection{$\Pi$ and $\Sigma$}
The computability structure of $\Pi$ and $\Sigma$ types follows a similar pattern of defining the cost algebraic structure $\Label{Alg}$ as $\F$ type.

\iblock{
    \mrow{\Pi\Label{Alg} : (A : \blue{\tpv}) \to (X : \blue{\tmvp}(A) \to \blue{\tpc}) \to \alg}
    \mrow{\Pi\Label{Alg}\proj{C} = \glue{e}{\red{\tmcp}(\red{\Pi}~A~X)}{\extTy{(a : \blue{\tmvp}(A)) \to \blue{\tmcp}(X~a)}{z}{\syn}{\red{\textsf{ap}}~e}}}
    \mhang{\Pi\Label{Alg}\proj{sstep}~c'~(e : \red{\tmcp}(\red{\Pi}~A~X), f : (a : \blue{\tmvp}(A)) \to \blue{\tmcp}(X~a)) =}{
        \mrow{\glueEx{\syn}{\red{\stepp}^{c'}(e)}{\lambda (a : \blue{\tmvp}(A)). \blue{\stepp}^{c'}(f~a)}}
    } 
}

To make sure the definition of $\Pi\Label{Alg}\proj{C}$ type-checks, we need to verify that under the syntactic phase:
\begin{align*}
    \syn \vdash & \\
    & \lambda (a : \blue{\tmvp}(A)). \blue{\stepp}^{c'}(f~a) \\
    =~& \lambda (a : \red{\tmvp}(A)). \red{\stepp}^{c'}(f~a) \tag*{under $\syn$}\\
    =~& \lambda (a : \red{\tmvp}(A)). \red{\stepp}^{c'}(\red{\app}~e~a) \tag*{by the extension type restriction}\\
    =~& \lambda (a : \red{\tmvp}(A)). \red{\app}~(\red{\stepp}^{c'}(e))~a \tag*{by $\red{\app_{\stepp}}$}\\
    =~& \red{\app}~(\red{\stepp}^{c'}(e)) \tag*{by $\eta$ of lambda}
\end{align*}

Again it is routine calculation to check that the associated equations hold, appealing to each equations $\stepp_0$, $\stepp_+$, and $\stepp_{\beh}$ in \red{red} and \blue{blue}.
\iblock{
    \mrow{\Pi\Label{Alg}\proj{sstep_{0}} = \checkmark}
    \mrow{\Pi\Label{Alg}\proj{sstep_{+}} = \checkmark}
    \mrow{\Pi\Label{Alg}\proj{sstep_{\beh}} = \eta_{\open_\syn}(\checkmark)}
    \mrow{}
    \mrow{\Label{\blue{\M}.}\blue{\Pi} : \extTy{(A : \blue{\tpv}) \to (\blue{\tmvp}(A) \to \blue{\tpc}) \to \blue{\tpc}}{z}{\syn}{\red{\Pi}}}
    \mrow{\Label{\blue{\M}.}\blue{\Pi}~A~X = \glueEx{\syn}{\red{\Pi}~A~X}{\Pi\Label{Alg}~A~X}}
    \mrow{}
    \mrow{\blue{\M}.\blue{\lambdap} : \extTy{((a : \blue{\tmvp}(A)) \to \blue{\tmcp}(X~a)) \to \blue{\tmcp}(\blue{\Pi}~A~X)}{z}{\syn}{\red{\lambdap}}}
    \mrow{\blue{\M}.\blue{\lambdap}~f = \glueEx{\syn}{\red{\lambdap}~f}{f}}
    \mrow{}
    \mrow{\blue{\M}.\blue{\app} : \extTy{\blue{\tmcp}(\blue{\Pi}~A~X) \to (a : \blue{\tmvp}(A)) \to \blue{\tmcp}(X~a)}{z}{\syn}{\red{\app}}}
    \mrow{\blue{\M}.\blue{\app}~e~a = (\pi_\bullet e)~a}
}

From this definition it is immediate that the $\beta$ and $\eta$ laws for $\Pi$ types hold:
\[
    \blue{\app}~(\blue{\lambdap}~f)~a = (\pi_\bullet(\glueEx{\syn}{\red{\lambdap}~f}{f}))~a = f~a 
\]
and 
\[
    \blue{\lambdap}~(\blue{\app}~e) = \glueEx{\syn}{\red{\lambdap}~(\red{\app}~e)}{\blue{\app}~e} = \glueEx{\syn}{e}{\pi_\bullet e} = e.
\]

The computability structure of $\Sigma$ types follows directly from \citet[\S 4.4.1.3]{sterling>thesis} since we are considering a purely value level $\Sigma$ type.

\subsubsection{Modalities}

The computability structure of the open modality, being a reader monad, displays a similar pattern as the $\Pi$ type as they are both encoded using the framework-level function space. That is, the semantics of object-level open modality $\Op^+$ is given by the meta-level open modality $\Op_{\beh}A \isdef \beh \to A$. 

\iblock{
    \mrow{\Label{\blue{\M}.\blue{\Op^+}} : \extTy{\blue{\tpv} \to \blue{\tpv}}{z}{\syn}{\red{\Op^+}}}
    \mrow{\Label{\blue{\M}.\blue{\Op^+}} A = \glueEx{\syn}{\red{\Op^+}A}{\glue{a}{\red{\tmvp}(\red{\Op^+}A)}{\extTy{\beh \to \blue{\tmvp}(A)}{z}{\syn}{\red{\app_{\Op}}~a}}}}
    \mrow{}
    \mrow{\Label{\blue{\M}.\blue{\lambdap_{\Op}}} : \extTy{(\beh \to \blue{\tmvp}(A)) \to \blue{\tmvp}(\blue{\Op^+}A)}{z}{\syn}{\red{\lambdap_{\Op}}}}
    \mrow{\Label{\blue{\M}.\blue{\lambdap_{\Op}}}~f = \glueEx{\syn}{\red{\lambdap_{\Op}}~f}{f}}
    \mrow{}
    \mrow{\Label{\blue{\M}.\blue{\app_{\Op}}} : \extTy{\blue{\tmvp}(\blue{\Op^+}A) \to \beh \to \blue{\tmvp}(A)}{z}{\syn}{\red{\app_{\Op}}}}
    \mrow{\Label{\blue{\M}.\blue{\app_{\Op}}}~a~z = (\pi_\bullet a)~z}
}

Similarly the semantics of object-level closed modality $\Cl^+$ is conveniently given by the closed modality $\Cl_{\beh}$ in the meta language of synthetic Tait computability.

\iblock{
    \mrow{\Label{\blue{\M}.\blue{\Cl^+}} : \extTy{\blue{\tpv} \to \blue{\tpv}}{z}{\syn}{\red{\Cl^+}}}
    \mhang{\Label{\blue{\M}.\blue{\Cl^+}} A = \glueEx{\syn}{\red{\Cl^+}A}{\glue{a}{\red{\tmvp}(\red{\Cl^+}A)}{\delta}}}{
        \mrow{\kw{where}}
        \mrow{\delta = \closed_{\syn}((ba : \closed_{\beh}(\blue{\tmvp}(A))) \times \try{ba}{\lambda(a' : \blue{\tmvp}(A)). \open_{\syn}(a = \red{\eta_{\Cl}}(a'))}{\beh}{\open_{\syn}(a = \red{\star}(\_))})}
    }
}

The two clauses of the try is identified under $\beh$ because of the $\red{\mathsf{law}_{\Cl}} : \red{\eta_{\Cl}}(a) = \red{\star}(z)$ equation. 

\iblock{
    \mrow{\Label{\blue{\M}.\blue{\eta_{\Cl}}} : \extTy{\blue{\tmvp}(A) \to \blue{\tmvp}(\blue{\Cl^+}(A))}{z}{\syn}{\red{\eta_{\Cl}}}}
    \mrow{\Label{\blue{\M}.\blue{\eta_{\Cl}}}~a = \glueEx{\syn}{\red{\eta_{\Cl}}~a}{\eta_{\closed_{\syn}}(\eta_{\closed_{\beh}}(a), \eta_{\open_{\syn}}(\checkmark : \red{\eta_{\Cl}}(a) = \red{\eta_{\Cl}}(a)))}}
    \mrow{}
    \mrow{\Label{\blue{\M}.\blue{\star}} : \extTy{\beh \to \blue{\tmvp}(\blue{\Cl^+}(A))}{z}{\syn}{\red{\star}}}
    \mrow{\Label{\blue{\M}.\blue{\star}}~z = \glueEx{\syn}{\red{\star}~z}{\eta_{\closed_{\syn}}(\star_{\beh}(z), \eta_{\open_{\syn}}(\checkmark : \red{\star}(z) = \red{\star}(\_))) }}
}

The equation $\blue{\mathsf{law}_{\Cl}}$ holds directly by the meta-level $\mathsf{law}$ for closed modality. The induction principle for closed modality is modeled by the try construct of meta-level closed modality.

\subsubsection{Equality types}
Since the extensional equality type $\textsf{eq}$ is a value type in \calf{}, it behaves no differently from the usual equality type in Martin-L\"of type theory. The computability structure of equality types is then exactly the same as one given in \citet[\S 4.4.1.4]{sterling>thesis}.

\subsubsection{Natural numbers}
As a final ingredient, we define the computability structure for the $\nat$ type, which is the subject of the canonicity result. This roughly follows \citet{buchholtz-schipp>2024}, adjusting to a call-by-push-value setting, and fixing a slight bug in that work.
\iblock{
    \mrow{\Label{\blue{\M}.\blue{\nat}} : \extTy{\blue{\tpv}}{z}{\syn}{\red{\nat}}}
    \mrow{\Label{\blue{\M}.\blue{\nat}} = \glueEx{\syn}{\red{\nat}}{\glue{e}{\red{\tmvp}(\red{\nat})}{\closed_{\syn}((n : \mathbb{N}) \times \open_{\syn}(e = \red{\sucp}^n(\red{\zero})))}}}
    \mrow{}
    \mrow{\Label{\blue{\M}.\blue{\zero}} : \extTy{\blue{\tmvp}(\blue{\nat})}{z}{\syn}{\red{\zero}}}
    \mrow{\Label{\blue{\M}.\blue{\zero}} = \glueEx{\syn}{\red{\zero}}{\eta_{\closed_{\syn}}(0, \eta_{\open_{\syn}}(\checkmark : \red{\zero} = \red{\zero}))}}
    \mrow{}
    \mrow{\Label{\blue{\M}.\blue{\sucp}} : \extTy{\blue{\tmvp}(\blue{\nat}) \to \blue{\tmvp}(\blue{\nat})}{z}{\syn}{\red{\sucp}}}
    \mhang{\Label{\blue{\M}.\blue{\sucp}}~e = \glueEx{\syn}{\red{\sucp}~e}{\mapp_{\closed_{\syn}}(\theta)(\pi_\bullet e)}}{
        \mrow{\kw{where}}
        \mrow{\theta : (n : \mathbb{N}) \times \open_{\syn}(\red{\sucp}^n(\red{\zero})) \to (n : \mathbb{N}) \times \open_{\syn}(\red{\sucp}^n(\red{\zero}))}
        \mrow{\theta~(n,p) = (1+n, \eta_{\open_{\syn}}(\checkmark : \red{\sucp}(e) = \red{\sucp}^{1+n}(\red{\zero})))}
    }
    \mrow{}
    \mhang{\Label{\blue{\M}.\blue{\indp}} : \{ (e : \blue{\tmvp}(\blue{\nat}))}{
        \mrow{\to (X : \blue{\tmvp}(\blue{\nat}) \to \blue{\tpc})}
        \mrow{\to \blue{\tmcp}(X~\blue{\zero})}
        \mrow{\to ((n : \blue{\tmvp}(\blue{\nat})) \to \blue{\tmcp}(X~n) \to \blue{\tmcp}(X~(\blue{\sucp}(n))))}
        \mrow{\to \blue{\tmcp}(X~e) \mid \syn \hookrightarrow \red{\indp} \} }
    }
    \mhang{\Label{\blue{\M}.\blue{\indp}}~e~X~e_0~e_1 = \try{\pi_\bullet e}{\lambda (n, p). \omega(n)}{\syn}{\red{\indp}~e~X~e_0~e_1}}{
        \mrow{\kw{where}}
        \mrow{\omega : (n : \mathbb{N}) \to \blue{\tmcp}(X~(\blue{\sucp}^n(\blue{\zero})))}
        \mrow{\omega (0) = e_0}
        \mrow{\omega (1+n) = e_1~(\blue{\sucp}^n(\blue{\zero}))~(\omega (n))}
    }
}

\begin{remark}
    In \citet{buchholtz-schipp>2024}, the definition of $\blue{\indp}$ is slightly erroneous in that in therein $p : \open_{\syn}(e = \red{\sucp}^n(\red{\zero}))$ goes into the meta-level induction $\omega$, but it is an equation about the term $e : \blue{\tmvp}(\blue{\nat})$ which never changes. The correct definition is as we have shown above. The role of proof $p$ here is to make sure $\omega(n)$ has the correct type $\blue{\tmcp}(X~e)$.
\end{remark}

\subsection{Fundamental theorem}\label{sec:fundamental}
We may finally conclude that the computability structure $\blue{\M}$ is a model of \calf{} in the sense that it satisfies the signature $\Sigma_{\calf{}}$. In particular, the computability structure $\blue{\M}$ restricts to the syntactic model $\red{\M}$ under the syntactic phase. Furthermore there are structure-preserving homomorphisms $\textsf{\red{blu}\blue{ify}} : \red{\M} \to \blue{\M}$ and $\eta_{\open_{\syn}} : \blue{\M} \to \red{\M}$ such that $\eta_{\open_{\syn}} \circ \textsf{\red{blu}\blue{ify}} = \textsf{Id}$:
\[\begin{tikzcd}[cramped,sep=small]
	{\textcolor{red}{\textsf{M}}} &&&& {\textcolor{blue}{\textsf{M}}} \\
	\\
	\\
	\\
	&&&& {\textcolor{red}{\textsf{M}}}
	\arrow["{\textsf{\textcolor{red}{blu}\textcolor{blue}{ify}}}", from=1-1, to=1-5]
	\arrow["{\textsf{Id}}"', from=1-1, to=5-5]
	\arrow["{\eta_{\open_{\syn}}}", from=1-5, to=5-5]
\end{tikzcd}\]
We recall our theorem statement, stating in the language of $\red{\M}$:
\begin{theorem}[Canonicity]
  For every term $e : \red{\tmcp}(\red{\F}(\red{\nat}))$, there exists $c : \costty$ and $n : \mathbb{N}$ such that $e = \red{\stepp}^c(\red{\retp}(\red{\sucp}^n(\red{\zero})))$.
\end{theorem}

\section{Interpreting the computability structure}\label{sec:interpretation}

The aforementioned theorem can be stated in the language of category of judgments $\mathcal{C}$:
\begin{theorem}[Canonicity]
    In the category of judgments of \calf{} $\mathcal{C}$, for every morphism $e : \red{\terminal_{\mathcal{C}}} \to \red{\tmcp}(\red{\F}(\red{\nat}))$, there exists $c : \costty$ and $n : \mathbb{N}$ such that there is a morphism $p : \red{\terminal_{\mathcal{C}}} \to e = \red{\stepp}^c(\red{\retp}(\red{\sucp}^n(\red{\zero})))$.
\end{theorem}

The proof strategy is as follows: we find a locally Cartesian closed category $\mathcal{E}$ and a locally Cartesian closed functor $\textsf{\red{blu}\blue{ify}} : \mathcal{C} \to \mathcal{E}$ according to the definition of $\blue{\M}$. The morphism $e : \red{\terminal_{\mathcal{C}}} \to \red{\tmcp}(\red{\F}(\red{\nat}))$ is then sent to a morphism $\textsf{\red{blu}\blue{ify}}(e) : \blue{\terminal_{\mathcal{E}}} \to \blue{\tmcp}(\blue{\F}(\blue{\nat}))$ in $\mathcal{E}$. The canonicity theorem is then proved by examining this morphism in $\mathcal{E}$.

The recipe of synthetic Tait computability is that the computability result can be constructed from a comma category $\mathcal{E} \isdef \presheaf{\mathcal{I}} \downarrow N_\rho$ where 
\iblock{
    \mrow{N_\rho : \mathcal{C} \to \presheaf{\mathcal{I}}}
    \mrow{N_\rho(X) = \homsetAlt{\mathcal{C}}{\rho-}{X}}
}
$N_\rho$ is a nerve functor from the category of judgments $\mathcal{C}$ introduced earlier to a chosen presheaf category $\presheaf{\mathcal{I}}$ along the shape $\rho : \mathcal{I} \to \mathcal{C}$. This category is called the \emph{Artin Gluing} of $N_\rho$. For example, in the case of the canonicity proof of ordinary Martin-L\"of type theory, the category $\mathcal{I}$ is chosen to be $\textbf{1}_{\textbf{Cat}}$ and shape $\rho : \textbf{1}_{\textbf{Cat}} \to \mathcal{C}$ is the terminal functor that sends everything to $\textbf{1}_{\mathcal{C}}$.

The semantics of \calf{} \cite[\S 5]{niu-sterling-grodin-harper>2022} (or more generally, semantics of phased type theory) is given by a two-world Kripke semantics, that is, presheaves on the “walking arrow” category $\mathcal{I} \isdef \{ \circ \to \bullet \}$, where $\circ$ is the behavioral world (the $\beh$avioral phase) and $\bullet$ is the cost-sensitive world (the ambient ``true'' phase). Therefore the nerve functor $N_\rho$ is given by the following:
\iblock{
    \mrow{\rho : \mathcal{I} \to \mathcal{C}}
    \mrow{\rho(\circ) = \red{\beh}}
    \mrow{\rho(\bullet) = \red{\terminal_{\mathcal{C}}}}
    \mrow{\rho(\circ \to \bullet) = ~! : \red{\beh} \to \red{\terminal_{\mathcal{C}}}}
}
The anti-monotonicity property of Kripke semantics, which says that a judgment holds under the ``true'' phase $\cdot \vdash A$ (as a morphism $\red{\terminal_{\mathcal{C}}} \to \red{A}$) implies it also holds under the behavioral phase $\beh \vdash A$ (as a morphism $\red{\beh} \to \red{A}$), is captured by the fact that pre-composing every morphism $a : \red{\terminal_{\mathcal{C}}} \to \red{A}$ with the unique morphism $! : \red{\beh} \to \red{\terminal_{\mathcal{C}}}$ gives a morphism $a ~\circ~ ! : \red{\beh} \to \red{A}$.

To be explicit, objects of the comma category $\mathcal{E} \isdef \presheaf{\mathcal{I}} \downarrow N_\rho$ are
\[
    \Biggl \langle \red{A} \in \mathcal{C}, \Presheaf[f]{S^{\bullet} \in \set}{S^\circ \in \set}, \Presheaff{g^\bullet : S^\bullet \to \homset{\red{\terminal_{\mathcal{C}}}}{\red{A}}}{g^\circ : S^\circ \to \homset{\red{\beh}}{\red{A}}} \Biggr \rangle
\]

such that the following diagram commutes:
\[\begin{tikzcd}[cramped,sep=small]
	{S^\bullet} &&& {\homset{\red{\terminal_{\mathcal{C}}}}{\red{A}}} \\
	\\
	\\
	{S^\circ} &&& {\homset{\red{\beh}}{\red{A}}}
	\arrow["{g^\bullet}", from=1-1, to=1-4]
	\arrow["f"', from=1-1, to=4-1]
	\arrow["{- \circ !}", from=1-4, to=4-4]
	\arrow["{g^\circ}", from=4-1, to=4-4]
\end{tikzcd}\]
where the morphism $-\circ !: \homset{\red{\terminal_{\mathcal{C}}}}{\red{A}} \to \homset{\red{\beh}}{\red{A}}$ is pre-composition with the unique morphism $! : \red{\beh} \to \red{\terminal_{\mathcal{C}}}$.

Informally an object should be understand as having a syntactic part $\red{A}$ from the category of judgments $\mathcal{C}$, and a semantic part $S^\bullet$ (representing proofs in the cost-sensitive phase) and $S^\circ$ (representing proofs in the behavioral phase) from the presheaf $\Psh{\mathcal{I}}$, and a pair of functions $g^\bullet$ and $g^\circ$ that makes sure that the syntax and semantics agree. 

Morphisms in this category are three tuples $\red{a} : \red{A} \to \red{B}$, $h^\bullet : S^\bullet \to T^\bullet$, and $h^\circ : S^\circ \to T^\circ$ such that the following squares commute: 
\[\begin{tikzcd}[cramped,sep=small]
	{S^\bullet} &&& {T^\bullet} & {S^\circ} &&&& {T^{\circ}} \\
	\\
	\\
	{\homset{\red{\terminal_{\mathcal{C}}}}{\red{A}}} &&& {\homset{\red{\terminal_{\mathcal{C}}}}{\red{B}}} & {\homset{\red{\beh}}{\red{A}}} &&&& {\homset{\red{\beh}}{\red{B}}}
	\arrow["{h^\bullet}", from=1-1, to=1-4]
	\arrow["{g^\bullet}"', from=1-1, to=4-1]
	\arrow["{k^\bullet}", from=1-4, to=4-4]
	\arrow["{h^{\circ}}", from=1-5, to=1-9]
	\arrow["{g^\circ}"', from=1-5, to=4-5]
	\arrow["{k^{\circ}}", from=1-9, to=4-9]
	\arrow["{\homset{\red{\terminal_{\mathcal{C}}}}{\red{a}}}"', from=4-1, to=4-4]
	\arrow["{\homset{\red{\beh}}{\red{a}}}"', from=4-5, to=4-9]
\end{tikzcd}\]

In particular there is a forgetful functor $\pi_\circ : \mathcal{E} \to \mathcal{C}$ that forgets the semantic part. Its action on objects is given by $\pi_\circ(\red{A}, S^\bullet, S^\circ, g^\bullet, g^\circ) = \red{A}$ and on morphisms by $\pi_\circ(\red{a}, h^\bullet, h^\circ) = \red{a}$.

Importantly, since the category of judgments $\mathcal{C}$ is locally cartesian closed, the Artin Gluing of a left-exact functor $N_\rho$ is also locally cartesian closed. This means that the internal language of $\mathcal{E}$ is a rich extensional dependent type theory, which is the language of synthetic Tait computability, that is capable of interpreting $\blue{\M}$. Normal dependent type theory constructs can be interpreted into arbitrary locally Cartesian closed categories \cite{seely>1984,hofmann>1994}, and we will show extra structures of synthetic Tait computability can be interpreted in $\mathcal{E}$ as well. What remains is to see what exactly is the morphism $\textsf{\red{blu}\blue{ify}}(e) : \blue{\terminal_{\mathcal{E}}} \to \blue{\tmcp}(\blue{\F}(\blue{\nat}))$ in this interpretation. 

\subsection{Interpreting phases and modalities}
\subsubsection{Phases} 
As phases are propositions, they are interpreted as distinguished subterminal objects. In particular, we have:
\begin{align*}
    \interp{\syn} &\isdef \Biggl \langle \red{\terminal_{\mathcal{C}}}, \Presheaf{\initial_{\set}}{\initial_{\set}}, \Presheaff{! : \initial_{\set} \to \homset{\red{\terminal_{\mathcal{C}}}}{\red{\terminal_{\mathcal{C}}}}=\terminal_{\set}}{! : \initial_\set \to \homset{\red{\beh}}{\red{\terminal_{\mathcal{C}}}}=\terminal_{\set}}  \Biggr \rangle \\ 
    \interp{\beh} &\isdef \Biggl \langle \red{\beh}, \Presheaf{\initial_{\set}}{\terminal_{\set}}, \Presheaff{! : \initial_{\set} \to \homset{\red{\terminal_{\mathcal{C}}}}{\red{\beh}}}{! : \terminal_{\set} \to \homset{\red{\beh}}{\red{\beh}}=\terminal_{\set}} \Biggr \rangle
\end{align*}

We can check that both $\interp{\syn}$ and $\interp{\beh}$ are indeed subterminal objects in $\mathcal{E}$.

The syntactic part $\pi_\circ$ of $\interp{\beh}$ is exactly $\red{\beh}$, validating \cref{ax:beh} that $\beh : \extTy{\univ_j}{z}{\syn}{\red{\beh}}$.

\subsubsection{Modalities}
The modalities $\open_{\syn}$ and $\closed_{\syn}$ are interpreted as the exponential functor $(-)^{\interp{\syn}}$ and pushout along projections of $A \times \interp{\syn}$ respectively. 

\begin{align*}
    \interp{\open_{\syn}} \Biggl \langle \red{A}, \Presheaf{S^\bullet}{S^\circ}, \Presheaff{g^{\bullet}}{g^\circ} \Biggr \rangle & \isdef \Biggl \langle \red{A}, \Presheaf{\homset{\red{\terminal_{\mathcal{C}}}}{\red{A}}}{\homset{\red{\terminal_{\mathcal{C}}}}{\red{A}}}, \Presheaff{\textsf{id} : \homset{\red{\terminal_{\mathcal{C}}}}{\red{A}} \to \homset{\red{\terminal_{\mathcal{C}}}}{\red{A}}}{- \circ ! : \homset{\red{\terminal_{\mathcal{C}}}}{\red{A}} \to \homset{\red{\beh}}{\red{A}}} \Biggr \rangle \\
    \interp{\closed_{\syn}} \Biggl \langle \red{A}, \Presheaf{S^\bullet}{S^\circ}, \Presheaff{g^{\bullet}}{g^\circ} \Biggr \rangle & \isdef \Biggl \langle \red{\terminal_{\mathcal{C}}}, \Presheaf{S^\bullet}{S^\circ}, \Presheaff{! : S^\bullet \to \homset{\red{\terminal_{\mathcal{C}}}}{\red{\terminal_{\mathcal{C}}}}=\terminal_{\set}}{! : S^\circ \to \homset{\red{\beh}}{\red{\terminal_{\mathcal{C}}}}=\terminal_{\set}} \Biggr \rangle
\end{align*}

In particular, we observe that any $\closed_\syn$-modal objects have trivial first and third components, and the only interesting part is the semantic presheaf $S^\bullet \to S^\circ$.

\subsubsection{Strict glue types}
To determine the interpretation of the strict glue type $\glue{x}{\red{A}}{B(x)}$, we observe that 
\begin{enumerate}
    \item $\red{A}$ is $\open_{\syn}$-modal, so its interpretation must be of the form $$\interp{\red{A}} = \Biggl \langle \red{A}, \Presheaf{\homset{\red{\terminal_{\mathcal{C}}}}{\red{A}}}{\homset{\red{\terminal_{\mathcal{C}}}}{\red{A}}}, \Presheaff{\textsf{id}}{- \circ !} \Biggr \rangle .$$
    \item For each closed term of type $\red{A}$, as a morphism $\red{a} : \red{\terminal_\mathcal{C}} \to \red{A}$, we have that $B(\red{a})$ is $\closed_\syn$-modal, so its interpretation must be of the form $$\interp{B(\red{a})} = \Biggl \langle \red{\terminal_{\mathcal{C}}}, \Presheaf{S_{\red{a}}^\bullet}{S_{\red{{a}}}^\circ}, \Presheaff{!}{!} \Biggr \rangle.$$
\end{enumerate}

Gluing them together, keeping in mind that strict glue type behaves like a $\Sigma$ type, we have:
\[
    \interp{\glue{x}{\red{A}}{B(x)}} = \Biggl \langle \red{A}, \Presheaf{
        \{ (\red{a}, b) \mid \red{a} \in \homset{\red{\terminal_{\mathcal{C}}}}{\red{A}}, b \in S_{\red{a}}^\bullet \}}{
        \{ (\red{a}, b) \mid \red{a} \in \homset{\red{\terminal_{\mathcal{C}}}}{\red{A}}, b \in S_{\red{a}}^\circ \}}, 
        \Presheaff{\pi_1}{(- \circ !)\circ \pi_1} \Biggr \rangle
\]
where 
$$(- \circ !) \circ \pi_1 : \{ (\red{a}, b) \mid \red{a} \in \homset{\red{\terminal_{\mathcal{C}}}}{\red{A}}, b \in S_{\red{a}}^\circ \} \to \homset{\red{\beh}}{\red{A}}$$ is the pre-composition of every $\red{a} \in \homset{\red{\terminal_{\mathcal{C}}}}{\red{A}}$ with the unique morphism $! : \red{\beh} \to \red{\terminal_{\mathcal{C}}}$.

\subsection{Interpreting other type structures}
As mentioned above, the eventual goal is to find the object corresponding to $\blue{\tmcp}(\blue{\F}(\blue{\nat}))$. We recall: 
\iblock{
    \mhang{\blue{\tmcp}(\blue{\F}(\blue{\nat})) = \glue{e}{\red{\tmcp}(\red{\F}(\red{\nat}))}{\closed_{\syn}((n : \mathbb{N}, bc : \closed_{\beh}\costty) \times \alpha)}}{
        \mrow{\alpha = \try{bc}{\lambda (c : \costty). \open_{\syn}(e = \red{\stepp}^c(\red{\retp}(\red{\sucp}^n(\red{\zero}))))}{\beh}{\open_{\syn}(e = \red{\retp}(\red{\sucp}^n(\red{\zero})))}}
    }
}

Notably the second component of $\blue{\tmcp}(\blue{\F}(\blue{\nat}))$ is $\closed_{\syn}$-modal, so its interpretation will have a trivial syntax part $\red{\terminal_{\mathcal{C}}}$; the only interesting part is the semantic presheaf, which we denote as $\interpAlt{-}$. Common types such as $\mathbb{N}$ and $\costty$ are interpreted as constant presheaves; $\Sigma$ types are interpreted in the usual way in a presheaf category. $\open_{\beh}$ erases the cost-sensitive part of the presheaf and $\closed_{\beh}$ trivializes the behavioral part of the presheaf. 

\begin{align*}
    \interpAlt{A} &\isdef \Presheaf{\interpAlt{A}^\bullet}{\interpAlt{A}^\circ} & 
    \interpAlt{\mathbb{N}} &\isdef \Presheaf[\text{id}]{\mathbb{N}}{\mathbb{N}} &
    \interpAlt{\costty} &\isdef \Presheaf[\text{id}]{\costty}{\costty} & \\
    \interpAlt{\beh} &\isdef \Presheaf{\initial_{\set}}{\terminal_{\set}} & 
    \interpAlt{\open_{\beh}A} &\isdef \Presheaf[\text{id}]{\interpAlt{A}^\circ}{\interpAlt{A}^\circ} &
    \interpAlt{\closed_{\beh}A} &\isdef \Presheaf[!]{\interpAlt{A}^\bullet}{\terminal_{\set}} &
\end{align*}

We have the interpretation of $\try{e}{f}{\beh}{g} : B$ as morphisms between presheaves:

\begin{multicols}{3}

    \[
    \interpAlt{e : \closed_{\beh} A} = 
    \begin{tikzcd}[cramped,sep=tiny]
        {\terminal_{\set}} &&& {\interpAlt{A}^\bullet} \\
        \\
        \\
        {\terminal_{\set}} &&& {\terminal_{\set}}
        \arrow["{e^{\bullet}}", from=1-1, to=1-4]
        \arrow[from=1-1, to=4-1]
        \arrow[from=1-4, to=4-4]
        \arrow[from=4-1, to=4-4]
    \end{tikzcd}
    \] 

    \[
    \interpAlt{f : A \to B} =
    \begin{tikzcd}[cramped,sep=tiny]
        {\interpAlt{A}^\bullet} &&& {\interpAlt{B}^\bullet} \\
        \\
        \\
        {\interpAlt{A}^\circ} &&& {\interpAlt{B}^\circ}
        \arrow["{f^\bullet}", from=1-1, to=1-4]
        \arrow[from=1-1, to=4-1]
        \arrow[from=1-4, to=4-4]
        \arrow[from=4-1, to=4-4]
    \end{tikzcd}\]

    \[
    \interpAlt{g : \beh \to B} = 
    \begin{tikzcd}[cramped,sep=tiny]
        {\initial_{\set}} &&& {\interpAlt{B}^\bullet} \\
        \\
        \\
        {\terminal_{\set}} &&& {\interpAlt{B}^\circ}
        \arrow[from=1-1, to=1-4]
        \arrow[from=1-1, to=4-1]
        \arrow[from=1-4, to=4-4]
        \arrow["{g^{\circ}}"', from=4-1, to=4-4]
    \end{tikzcd}
    \]
\end{multicols}

\[
\interpAlt{\try{e}{f}{\beh}{g} : B} \isdef
\begin{tikzcd}[cramped,sep=small]
	{\terminal_{\set}} &&& {\interpAlt{B}^\bullet} \\
	\\
	\\
	{\terminal_{\set}} &&& {\interpAlt{B}^\circ}
	\arrow["{f^{\bullet}~\circ~ e^{\bullet}}", from=1-1, to=1-4]
	\arrow[from=1-1, to=4-1]
	\arrow[from=1-4, to=4-4]
	\arrow["{g^\circ}"', from=4-1, to=4-4]
\end{tikzcd}\]

Lastly, we have the interpretation of $\open_{\syn}(\red{a} = \red{b})$ as:
\[
    \interpAlt{\open_{\syn}(\red{a} = \red{b})} = \Presheaf{\homset{\red{\terminal_{\mathcal{C}}}}{\red{a} = \red{b}}}{\homset{\red{\beh}}{\red{{a}} = \red{{b}}}}
\]

From here we can conclude that the interpretation of $\blue{\tmcp}(\blue{\F}(\blue{\nat}))$ is:
\[
    \interpAlt{\blue{\tmcp}(\blue{\F}(\blue{\nat}))} = \Presheaf{
        \{(e, p) \mid e \in \homset{\red{\terminal_{\mathcal{C}}}}{\red{\tmcp}(\red{\F}(\red{\nat})), n \in \mathbb{N}, c \in \costty, p \in \homset{\red{\terminal_{\mathcal{C}}}}{e = \red{\stepp}^c(\red{\retp}(\red{\sucp}^n(\red{\zero})))}} \}
    }{
        \{(e, p) \mid e \in \homset{\red{\terminal_{\mathcal{C}}}}{\red{\tmcp}(\red{\F}(\red{\nat})), n \in \mathbb{N}, c \in \costty, p \in \homset{\red{\beh}}{e = \red{\retp}(\red{\sucp}^n(\red{\zero}))}} \}
    }
\]

and hence

\[
    \interp{\blue{\tmcp}(\blue{\F}(\blue{\nat}))} = \Biggl \langle \red{\tmcp}(\red{\F}(\red{\nat})), 
    \Presheaf{
        \interpAlt{\blue{\tmcp}(\blue{\F}(\blue{\nat}))}^\bullet
    }{
        \interpAlt{\blue{\tmcp}(\blue{\F}(\blue{\nat}))}^\circ
    },
    \Presheaff{\pi_1}{(- \circ !)\circ \pi_1}
    \Biggr \rangle
\]

\subsection{Fundamental theorem}

We can now transpose the fundamental theorem stated in the language of dependent type theory in \cref{sec:fundamental} to the language of category theory. In particular, we have:
\begin{enumerate}
    \item The model $\blue{\M}$ of $\Sigma_{\calf{}}$ stated in the internal language of $\mathcal{E}$ induces a unique locally Cartesian closed functor $\textsf{\red{blu}\blue{ify}} : \mathcal{C} \to \mathcal{E}$.
    \item The construction $\open_{\syn}(\blue{\M} = \red{\M})$ means that the syntactic part of any $\textsf{\red{blu}\blue{ify}}(A)$ is still $A$. In other words, $\textsf{\red{blu}\blue{ify}}$ is a section of $\pi_\circ$ in the sense that $\pi_\circ \circ \textsf{\red{blu}\blue{ify}} = \textsf{Id}_{\mathcal{C}}$.
\end{enumerate}
This is illustrated in the following diagram:
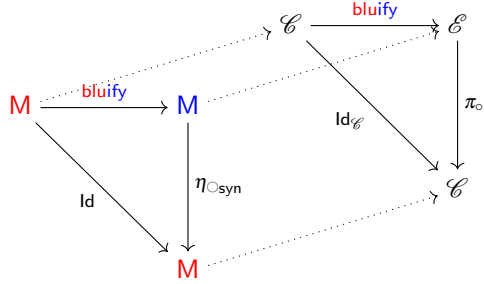
\[\begin{tikzcd}[cramped,sep=small]
	&&&&&& {\mathcal{C}} &&&& {\mathcal{E}} \\
	\\
	{\textcolor{red}{\textsf{M}}} &&&& {\textcolor{blue}{\textsf{M}}} \\
	\\
	&&&&&&&&&& {\mathcal{C}} \\
	\\
	&&&& {\textcolor{red}{\textsf{M}}}
	\arrow["{\textsf{\textcolor{red}{blu}\textcolor{blue}{ify}}}", from=1-7, to=1-11]
	\arrow["{\textsf{Id}_{\mathcal{C}}}"', from=1-7, to=5-11]
	\arrow["{\pi_\circ}", from=1-11, to=5-11]
	\arrow[dotted, from=3-1, to=1-7]
	\arrow["{\textsf{\textcolor{red}{blu}\textcolor{blue}{ify}}}", from=3-1, to=3-5]
	\arrow["{\textsf{Id}}"', from=3-1, to=7-5]
	\arrow[dotted, from=3-5, to=1-11]
	\arrow["{\eta_{\open_{\syn}}}", from=3-5, to=7-5]
	\arrow[dotted, from=7-5, to=5-11]
\end{tikzcd}\]

\subsection{Read off the canonicity theorem}

Suppose we have a morphism $e : \red{\terminal_{\mathcal{C}}} \to \red{\tmcp}(\red{\F}(\red{\nat}))$. Lifting it via functor $\textsf{\red{blu}\blue{ify}}$, we have a corresponding morphism $\textsf{\red{blu}\blue{ify}}(e) : \blue{\terminal_{\mathcal{E}}} \to \blue{\tmcp}(\blue{\F}(\blue{\nat}))$ in the category $\mathcal{E}$. That is, a morphism from 
 
\[
    \Biggl \langle 
        \red{\terminal_{\mathcal{C}}}, 
        \Presheaf{\terminal_{\set}}{\terminal_{\set}},
        \Presheaff{!}{!}
    \Biggr \rangle
    \qquad \text{ to } \qquad 
    \Biggl \langle \red{\tmcp}(\red{\F}(\red{\nat})), 
    \Presheaf{
        \interpAlt{\blue{\tmcp}(\blue{\F}(\blue{\nat}))}^\bullet
    }{
        \interpAlt{\blue{\tmcp}(\blue{\F}(\blue{\nat}))}^\circ
    },
    \Presheaff{\pi_1}{(- \circ !)\circ \pi_1}
    \Biggr \rangle.
\]

Examining this morphism, it consists of (among others) $e' : \red{\terminal_{\mathcal{C}}} \to \red{\tmcp}(\red{\F}(\red{\nat}))$ and $h^\bullet : \terminal_\set \to \interpAlt{\blue{\tmcp}(\blue{\F}(\blue{\nat}))}^\bullet$ such that the following diagram commutes:

\[\begin{tikzcd}[cramped,sep=small]
	{\terminal_\set} &&& {\interpAlt{\blue{\tmcp}(\blue{\F}(\blue{\nat}))}^\bullet} \\
	\\
	\\
	\homset{\red{\terminal_{\mathcal{C}}}}{\red{\terminal_{\mathcal{C}}}} = \terminal_\set &&& {\homset{\red{\terminal_{\mathcal{C}}}}{\red{\tmcp}(\red{\F}(\red{\nat}))}}
	\arrow["{h^\bullet}", from=1-1, to=1-4]
	\arrow[from=1-1, to=4-1]
	\arrow["{\pi_1}", from=1-4, to=4-4]
	\arrow["{e'}"', from=4-1, to=4-4]
\end{tikzcd}\]

Taking $\pi_\circ (\textsf{\red{blu}\blue{ify}}(e) ) = e'$. By the fundamental theorem, $\pi_\circ \circ \textsf{\red{blu}\blue{ify}}$ must be the identity, so $e' = e$. Therefore we have:
\[\begin{tikzcd}[cramped,sep=small]
	{\terminal_\set} &&& {\interpAlt{\blue{\tmcp}(\blue{\F}(\blue{\nat}))}^\bullet} \\
	\\
	\\
	\terminal_\set &&& {\homset{\red{\terminal_{\mathcal{C}}}}{\red{\tmcp}(\red{\F}(\red{\nat}))}}
	\arrow["{h^\bullet}", from=1-1, to=1-4]
	\arrow[from=1-1, to=4-1]
	\arrow["{\pi_1}", from=1-4, to=4-4]
	\arrow["{e}"', from=4-1, to=4-4]
\end{tikzcd}\]

Now this diagram tells us that $$h^{\bullet} \in \{(e, p) \mid e \in \homset{\red{\terminal_{\mathcal{C}}}}{\red{\tmcp}(\red{\F}(\red{\nat})), n \in \mathbb{N}, c \in \costty, p \in \homset{\red{\terminal_{\mathcal{C}}}}{e = \red{\stepp}^c(\red{\retp}(\red{\sucp}^n(\red{\zero})))}} \}$$ and $\pi_1(h^\bullet) = e$. Therefore $$\pi_2(h^\bullet) : \red{\terminal_{\mathcal{C}}} \to e = \red{\stepp}^c(\red{\retp}(\red{\sucp}^n(\red{\zero})))$$ is exactly a proof that $e =  \red{\stepp}^c(\red{\retp}(\red{\sucp}^n(\red{\zero})))$ for some $n \in \mathbb{N}$ and $c \in \costty$. 

$\qed$

\section{Conclusion and future work}\label{sec:conclusion}

In this work, we proved the canonicity property of \calf{} using the method of synthetic Tait computability pioneered by \citet{sterling>thesis}. The core idea is that meta-theoretic properties of a type theory can be proved in a sufficiently rich glued category, whose internal language is an extensional dependent type theory enriched with the syntax/semantics phase distinction. Then the construction of a model of the object type theory in the glued category can be realized by a programming exercise in the internal language. The use of logical framework and synthetic phase distinctions gives us a succinct, modular proof of logical relations.

\subsection{Related work}

\subsubsection{Cost-aware logical framework}
The reason why it is worth studying the metatheory of \calf{} as presented in this work is that \calf{} is already a fruitful research direction to study mechanized cost analysis of functional and amortized algorithms and data structures \cite{grodin-harper>2024,li-grodin-harper>2023,grodin-li-harper>2025,grodin-niu-sterling-harper>2024} and cost-aware denotational semantics \cite{niu-harper>2023,niu-sterling-harper>2024}. The canonicity property proved in this work is an important (albeit basic) meta-theoretic property that ensures the soundness of the cost analysis. A more comprehensive overview of the cost-aware logical framework can be found in Niu's thesis \cite{niu>thesis}.

\subsubsection{Traditional logical relations}

Meta-theoretic properties of type theories are traditionally proved using logical relations, also known as Tait's computability method \cite{tait>1967,plotkin>1973}. Here we draw a comparison between this work and pedagogical materials on traditional logical relations of a cost-aware phased type theory \cite{harper_cost>2025} and call-by-push-value \cite{harper_cbpv>2025}. \citet{harper_cost>2025} considers a lax type theory and logical relations over an operational semantics, but is otherwise exactly the same as \calf{} (step counting as an effect and a behavioral phase that synthetically erases the cost). Many proof burdens in traditional logical relations such as anti-monotonicity of the Kripke logical relations, head-expansion lemmas, and the usual tedious work on substitutions are silently discharged in the method of synthetic Tait computability as shown in this work, highlighting the advantages of the synthetic approach.

\subsection{Future work}

Here we discuss some possible extensions of \calf{}, whose meta-theoretic properties should be able to be proved using synthetic Tait computability.

\subsubsection{Universes in \calf{}}

We naturally expect \calf{}, as a dependent type theory, to have a universe hierarchy. The difficulty lies in the polarized setting of call-by-push-value: the universes of value types ought to be themselves value types, but what about the universes of computation types? In two independent prior extension of \calf{}, both options are explored. In the work of \citet{niu-harper>2023}, following \citet{pedrot-tabareau>2019}, the universe of computation types is a computation type; this allows effects to be tracked on the type level: for example, a term of type $\mstep{c}{\F{\nat}}$ must at least incur $c$ cost and return a natural number. In the work of \citet{grodin-li-harper>2025}, following \citet{vakar>2015,vakar>thesis}, the universe of computation types is a value type; this way they can have computation types such as $\F{\listty{\nat}}$ as the implementation type of data structures. A priori it is not clear which option is ``correct''; nevertheless either options should be able to be modeled in the synthetic Tait computability following Sterling's treatment of universes in dependent type theory \cite[\S 4.4.3]{sterling>thesis}.

\subsubsection{Other effects in \calf{}}

The presentation of \calf{} in this work is a simple language with only cost effects. It is natural to extend it with other effects. \citet{grodin-niu-sterling-harper>2024} explored effects such as state and probabilistic sampling in the context of \calf{} by considering an intrinsic ordering on programs $e \le e'$ that expresses that $e$ and $e'$ are equivalent in terms of other effects but $e$ may incur less cost. We expect to be able to model these extensions in the synthetic Tait computability by considering a more complicated glued category $\mathcal{E}$ following the semantics of \citet{grodin-niu-sterling-harper>2024}. Other effects such as partiality should likewise be able to be modeled following the semantics of \citet{niu-sterling-harper>2024}.

\section*{Acknowledgements}
We acknowledge Harrison Grodin for valuable discussions that directly and indirectly inspired this work. This material is based upon work supported by the United States Air Force Office of Scientific Research under grant number FA9550-21-0009 and FA9550-23-1-0434 (Tristan Nguyen, program manager). Any opinions, findings and conclusions or recommendations expressed in this material are those of the authors and do not necessarily reflect the views of the AFOSR.

\bibliographystyle{plainnat}
\bibliography{main}

\end{document}